\documentclass[journal]{IEEEtran}

\usepackage[T1]{fontenc}
\usepackage{xcolor}
\usepackage{amsmath}
\usepackage{graphicx}
\usepackage{epstopdf}
\usepackage{multicol}
\usepackage{multirow}
\usepackage{xspace}
\usepackage{comment}
\usepackage{textcomp}

\usepackage[abs]{overpic} 

\graphicspath{{figs/}}
\newcommand{\fig}[1]{Fig.~\ref{#1}}

\begin{document}

\title{A Square Peg in a Round Hole: The Complex Path for Wireless in the Manufacturing Industry}

\author{
    Borja~Martinez,~\IEEEmembership{Senior~Member,~IEEE,}
    Cristina~Cano,
    Xavier~Vilajosana,~\IEEEmembership{Senior~Member,~IEEE}%
    \thanks{B.\,Martinez and C.\,Cano are with IN3 at Universitat Oberta de Catalunya.}
    \thanks{X.\,Vilajosana is with IN3 at Universitat Oberta de Catalunya and Worldsensing S.L.}%
}


\maketitle
\IEEEpeerreviewmaketitle

\begin{abstract}
The manufacturing industry is at the edge of the 4th industrial revolution,
a paradigm of integrated architectures in which the entire production chain (composed of machines, workers and products) is intrinsically connected.
Wireless technologies can add further value in this manufacturing revolution.
However, we identify some signs that indicate that wireless technology could be left out 
of the next generation of smart factory equipment.
This is particularly relevant considering that the heavy machinery characteristic of this sector can last for decades.
We argue that at the core of this issue there is a mismatch between industrial needs and the interests of academic and partly-academic sectors (such as standardization bodies).
We base our claims on surveys from renowned advisory firms and interviews with industrial actors, which we contrast with results from content analysis of scientific articles.
Finally, we propose some convergence paths that, while still retaining the degree of novelty required for academic purposes, are more aligned with industrial concerns.
\end{abstract}


\section{Introduction}
\label{sec:intro}

\IEEEPARstart{T}{he} success of wireless communication is unquestionable today. 
Wireless technologies have become a commodity in our society,
in which ubiquitous connectivity of mobile gadgets, 
wearables and home appliances is the norm.
Connection to wireless networks is now possible in a range of scenarios, including urban, rural, indoor 
and transportation systems.
Some industries have also integrated wireless communications in their operations. 
To name a few, wireless is now present in critical infrastructure monitoring, logistics, traffic management, utility metering and healthcare solutions. 
However, despite of the efforts to provide industrial wireless solutions, some sectors seem reluctant 
to widespread adoption.
This is the case, to a large extent, for the manufacturing industry, including the manufacturing process, electronics, aerospace, automotive and machine tool sectors.

\begin{figure}[t!]
  \centering
  \includegraphics[width=1.0\columnwidth]{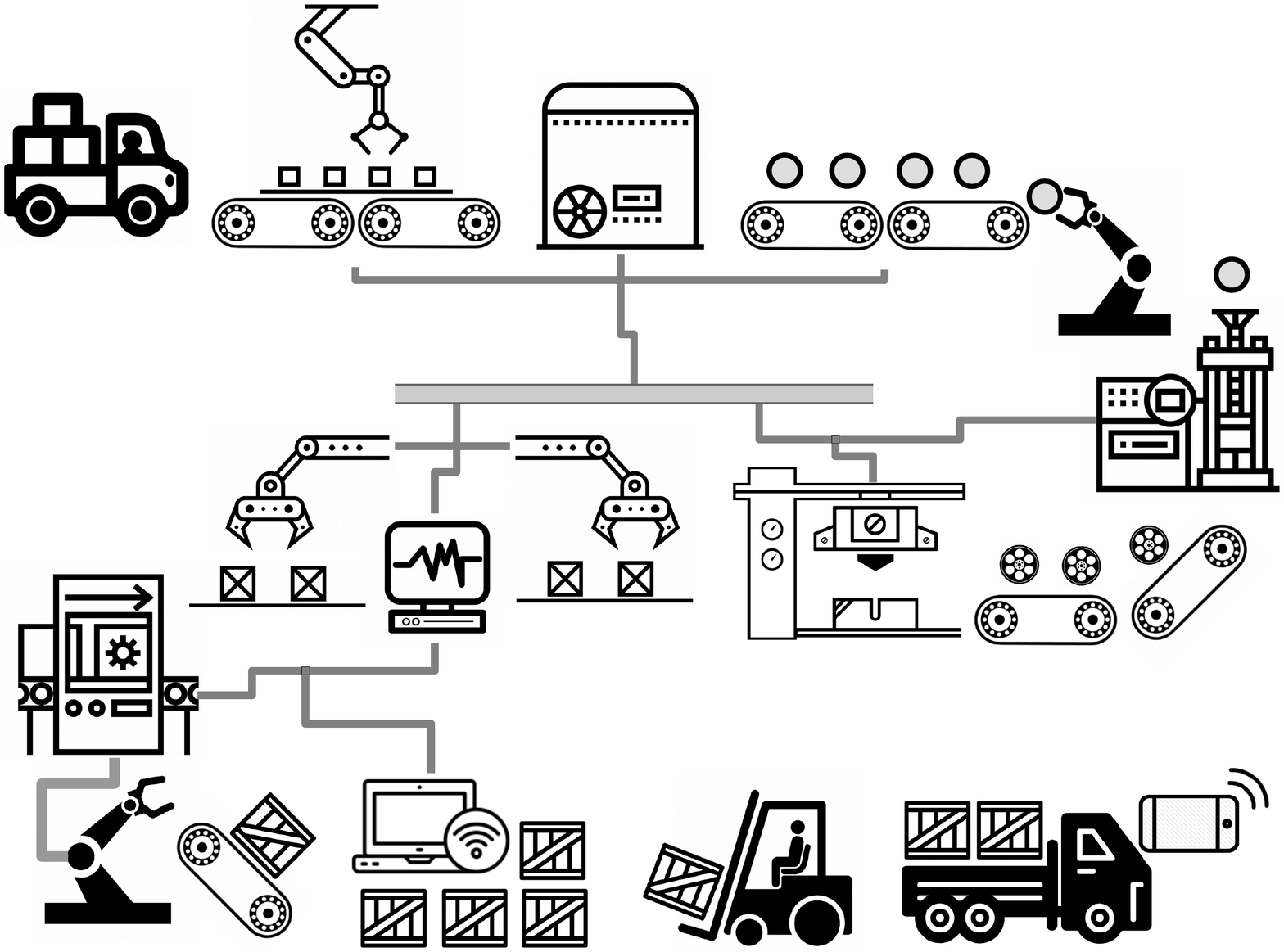}
  \caption{Interconnection of different entities of the manufacturing process following the Industry 4.0 vision.}
  \label{fig:smart-factory}
\end{figure}

The lack of massive adoption of wireless technologies by the manufacturing industry should be, at least, somewhat surprising to academics.
The efforts over recent years dedicated to transforming wireless technologies into suitable industrial solutions have been huge \cite{Huang2018}. 
The term coined as the Industrial Internet of Things (IIoT), 
which should enable the hyper-connected vision of the Industry 4.0 (\fig{fig:smart-factory}), 
has indeed gathered the interest of many researchers.
Plenty of solutions have been designed to cope with industrial requirements, or, more precisely, with what academics 
believe manufacturing industrial requirements are.
Despite these efforts,
it is not by any means clear when this massive adoption will occur and, more importantly, what are the reasons for this apparent delay.

It seems that the industry is indeed aware of wireless potential benefits.
Wireless has succeeded in other industrial sectors (such as chemical, oil and gas).
In these factories, most of the flow metering equipment used the Highway Addressable Remote Transducer (HART) protocol, a fieldbus protocol designed in the 80s. 
Later, in the 2000s, a wireless extension of HART was promoted by a consortium of 37 leading industries in order to bring wireless connectivity while keeping protocol features and semantics. 
WirelessHART became a natural evolution that succeeded in these factories. 
Thus, if the technology seems ready and the industry is aware of its potential benefits, the question arises:
Why is wireless not yet succeeding in the manufacturing sector?

This article attempts to gain insight into that question with special focus on the role of academia and partly-academic actors (such as standardization bodies).
In particular, we:
\begin{itemize}
\item Use different indicators to assess the penetration of wireless technologies in the manufacturing industry.
Direct evaluation is beyond our means, so we rely on opinion surveys conducted by renowned advisory firms
(such us Gartner and Morgan\&Stanley), interviews with industrial representatives and the inspection of machinery portfolios.
 \item Point out the potential causes preventing wireless adoption by this sector based on the information gathered.
 \item Carry out an honest self-reflection exercise to analyze to which extent academia contributes to this lack of success. 
 As a result, we identify and examine a mismatch between research directions and industrial reality.
 \item Devise academic-industrial convergence paths to move forward, hoping to contribute to the materialization of a wireless industrial revolution in the upcoming years.
\end{itemize}

{Note that this article does not aim to analyze industrial barriers 
related to internal organization, accountability, upfront investments or functional strata. 
It also does not aim to delve into other potential barriers such as the lack of predisposition 
of the manufacturing sector towards promoting research activities. 
Although probably equally important to the barriers discussed in this article, 
these other obstacles are beyond both our expertise and the scope of academia.}

\section{A Reality Check}
\label{sec:reality}

In this section we analyze the status of the IIoT and assess the level 
to which wireless has been adopted by
the manufacturing industry based on indicators such as surveys from advisory firms, interviews with industrial actors and portfolio inspection.

\subsection{Predictions and status of the IIoT}

In the 1990s, the Internet revolution redrew the Business-to-Consumer 21st century's sectors 
such as the media, retail and financial services. 
Likewise, the IoT is destined to completely redefine other sectors such as manufacturing, energy, agriculture and transportation \cite{WEF2015}. 
Although the {change} seems imminent, the rate of development may not be homogeneous in these different sectors.

In the year 2011, in a famous white paper that is still widely cited today, Cisco predicted 50 billion connected devices by 2020 \cite{Cisco2011}. 
Nowadays, approaching 2020, the forecasts are more conservative, calculated according to different sources between 20 and 30 billion \cite{Mumtaz2017}. 
One of the reasons for this downward revision can be attributed to more conservative sectors, among which the manufacturing industry stands out.

The manufacturing industry is expected to evolve towards a
distributed organization of production with connected products,
equipment, processes and logistics \cite{Wollschlaeger2017}.
These new interactions may result in unprecedented
levels of productivity and operational efficiency. 
Companies, however, are still struggling to understand this conceptual step and, above all,
to demonstrate how this concept can bring value to
their operations \cite{WEF2015}. Even though there is unquestionable
interest in adopting new solutions \cite{HungGartner2017}, most enterprises do not
know what to do with the broad spectrum of new technologies. This fact, along with other important barriers
such as concerns about cyber-security, interoperability and upfront investment \cite{McKinsey2015};
are slowing down adoption, at least considering the
most optimistic forecasts.

A recent report from Verizon \cite{Verizon2017} indicates that,
although the IoT is still not part of the manufacturing process, 
interest in it is high enough to expect high levels of adoption in the years to come, mainly to improve operational efficiency and productivity.
This is precisely why the manufacturing industry is considered the sector with the greatest growth potential \cite{McKinsey2015}.
With high probability, we may be at a turning point.

\subsection{Indicators of wireless adoption} 

Building upon the conclusions above, we now aim to get more insight into the current level of penetration of wireless in this industry. 
As a first step, we look at the penetration of Industry 4.0 in the manufacturing sector,
as wireless technologies are considered to be among its technological enablers.
Infosys published a survey of more than 400 manufacturing companies across 5 regions (China, France, Germany, UK, US) conducted in 2015 \cite{Infosys2015}. 
The results showed that the penetration of Industry 4.0 in this sector was low. 
That is, only 15\% of the surveyed companies have implemented \emph{dedicated strategies} for asset efficiency. 
Industry 4.0 may or may not include wireless technologies, 
so we can infer from this report that the penetration of wireless technologies may be even lower. 	

Despite having some limitations, another way to assess wireless adoption is to perform targeted interviews with prominent actors. 
In the second half of 2017 {interviews} were carried out with the maintenance and engineering teams of pioneering manufacturing industries in the automotive, pharmaceutical (blistering), machinery and industrial robotics sectors \cite{mmtc2017}. 
These interviews revealed that none of these industries had wireless technologies integrated in their processes. 

Another indicator is the availability of products with wireless I/O modules in the portfolios of industrial automation companies. 
We surveyed the portfolios of Rockwell Automation, ABB, Emerson, Schneider Electric, Honeywell and Mitsubishi Electric. 
We observed that wireless modules are not typically integrated in their products. 
Indeed, with the only exception of flow meters and sensors, we did not find any industrial equipment 
with integrated wireless 
(the common practice is to offer external wireless modules that can be attached to their products). 
While the absence of wireless in native equipment does not provide a quantitative measure of the degree of wireless penetration, it is indeed an indicator that the use is not widespread. Otherwise we would expect the offer of machinery with integrated wireless to be the norm.


\section{Wireless Shortcomings}
\label{sec:causes}

\vspace{5mm}

One of the main obstacles identified in our interviews is the poor perception of wireless technologies, especially regarding reliability \cite{mmtc2017}. 
Moreover, industrialists are particularly reluctant to changing something that is already working reasonably well.
This can be seen as a manifestation of resistance to change, 
but it is also a rule derived from the industrialist's experience. 
The adoption of wireless in the manufacturing process may introduce new problems alien to the technologies being replaced.
For example, one of the most obvious advantages of wireless is its lower installation cost. 
However, in many companies, the cost of stopping a production line due to a failure in communications can be much higher than the cost of wiring during installation (which occurs at scheduled stops).
{Thus, we must be aware of the current limitations of wireless technologies that are relevant to the industry so that we can devise ways forward.}
We review these next.

\subsection{Performance \& reliability}
No one will adopt a new technology that does not offer, at least, similar features as the one being replaced. 
The new technology must be mature enough for the replacement to run smoothly. 
Although bandwidth may not be an issue (in many factories communication systems are over-dimensioned), 
it is still not straightforward for wireless systems {to reach the level of reliability required 
by the manufacturing industry. This sector, indeed, is especially sensitive to reliability,  
as commercial margins are minimal and therefore production lines are extremely optimized.}

The attempts to increase performance and reliability in the WiFi arena have been focused on eliminating collisions but either the solutions are not available in commercial cards (such as PCF and HCCA) or they do not meet the expectations \cite{tramarin2018calibrated}.
In the IoT, common techniques are to allocate resources deterministically and perform frequency-hopping to deal with the unreliable nature of the wireless medium.
However, being optimized under a low-energy constraint, these protocols provide very limited bandwidth.
{Other more suitable technologies, such as 5G ultra-reliable low-latency communications or mmWave are either still in their infancy or have not yet been considered for these use cases.}

\subsection{Obsolescence \& technology cycles}

When a technology becomes obsolete, change is unavoidable. 
However, the life cycles of some industrial technologies seem to be never-ending. 
Some protocols still in use today, such as the RS485, date from the seventies. 
But when the time for change comes, there is the perception (actually well-founded) 
that obsolescence, the life cycles of wireless, are much shorter than that of wired technologies. 
The constant emergence of new technologies and their vertiginous evolution,
which could be seen as an advantage, 
is a handicap from an industrial perspective. 
Indeed, the adoption of wireless technologies generates uncertainty in aspects 
as important as long-term maintenance and technical support.
An example that illustrates this uncertainty is ZigBee,
which after 15 years since being standardized and several revisions, 
it has not consolidated.
Meanwhile, many other competitive alternatives have appeared on the market.

\subsection{Fragmentation}
There is a perception of a fragmented wireless market. 
Fragmentation provides opportunities but at the same time introduces uncertainty to non-expert adopters. 
Aside from the complexity of decision-making,
industrialists ask themselves how a technology will evolve and, above all,  
who is supporting this evolution. 
The current wireless landscape is not favorable. 
Multiple technologies have been standardized and many proprietary alternatives are constantly being offered \cite{Huang2018}. 
To name a few, we can mention in the wireless short-range and mesh networks: 
DECT-ULE, WirelessHART, ISA100.11a, 6TiSCH, Zigbee, Z-Wave, Thread, WiFi, BLE/Mesh \cite{Li2017}. In the long range space we can find Wireless M-Bus, LoRaWAN, Sigfox, Weightless, Igenu, DASH7, WISUN and NB-IoT, amongst others \cite{Raza2017}.

In turn, standardization committees are isolated in the creation of their own communication protocols and these are rarely designed for interoperability, particularly if they compete for a dominant position. 
As an example, there is a lack of interoperability of wireless protocols such as ISA100.11a, WirelessHART and 6TiSCH, all of them addressing similar scenarios and using the same physical and MAC layers. 

\subsection{Security}

Wireless technologies undoubtedly introduce security concerns for the industry.
Typical connections in plant floors are wired, 
which means that there is a physical barrier for a potential attacker.
In contrast, wireless systems are prone to attacks
from outside the factory premises.
While there has been considerable research efforts dedicated to secure wireless communications, the perception from the industry is that wireless may expose their systems in some way.  
The cost of a security attack is too high for an industrialist to assume
, especially when there are alternative wired solutions available that intrinsically minimize security issues.

\section{The Industry-Academia Mismatch}
\label{sec:mismatch}

Research can be a fundamental enabler
for overcoming the limitations presented above.
However, we have identified an imbalance between industrial needs and academic interests.
In this section we analyze this mismatch as a self-reflection exercise. 

\begin{figure*}[ht]
  \centering
  \includegraphics[width=1.0\textwidth]{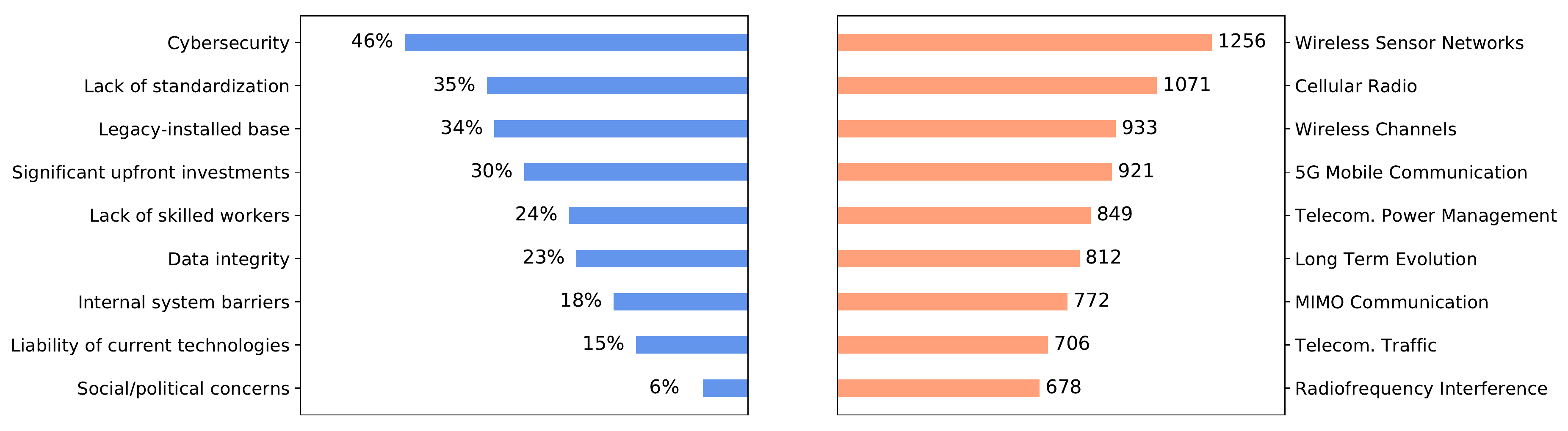}
  \caption{(Left) The most important industrial barriers to IIoT adoption 
  (percentage of surveyed companies that indicate the items listed as barriers) according to \cite{Morgan2016} vs. 
  (Right) the most used keywords in wireless research (paper count among the selected papers in our corpus).}
  \label{fig:iiot-morgan}
\end{figure*}

In order to support our statements throughout this section, 
we used content-analysis techniques applied to scientific articles.
In particular, we analyzed a corpus of more than 20,000 contributions to distinguished scientific journals and conferences. 
Specifically, this corpus, obtained through IEEEXplore, consists of all the articles published in 2015, 2016 and 2017
in the following IEEE journals: Communication Letters, Communications Magazine, Sensors Journal, Transactions on Industrial Informatics, Transactions on Mobile Computing, Transactions on Wireless Communications and Wireless Communication Letters;
as well as IEEE conferences (including their corresponding workshops):  Globecom, ICC, Infocom, Sensors and WCNC. 
 
From these 20,000 contributions, we then filtered the papers in which any of the following terms related to wireless appear in the abstract: 
\emph{wireless, wifi, mmwave, mmw, wsn, lte, rfid, zigbee, bluetooth, vanet, 4g, 5g, LoRa, Sigfox, lpwa, tsch}.
This filtering resulted in 9,011 papers, 
which are the ones that we use in the following analysis.
From these articles we analyzed both their INSPEC Controlled Terms and the contents of the abstract.\footnote{Code and corpus are available at: https://bitbucket.org/wineuoc/wireless-iot-landscape-ieeexplore-dataset}
The INSPEC Terms are used to extract conclusions on what are 
the most popular topics according to the academia,
while the contents of the abstracts allow to check the frequency 
of certain words we associate with some topics.
While the analysis of keywords is more generic, 
the abstracts provide a deeper understanding of academic interests.

\subsection{Macroscopic view}

To gain more insight into the most relevant academic interests, we analyzed the frequency of the INSPEC Terms in our corpus.
{In \fig{fig:iiot-morgan}} we compare the identified barriers for IIoT adoption 
that the industry reported in 
\cite{Morgan2016} with the 9 most frequent terms.
We fail to see any evident manifestation of the barriers perceived by the industry in the most {popular} interests of academia.
For instance, security and standardization are not among the major academic interests.

\begin{figure*}[ht]
  \centering
  \includegraphics[width=0.9\textwidth]{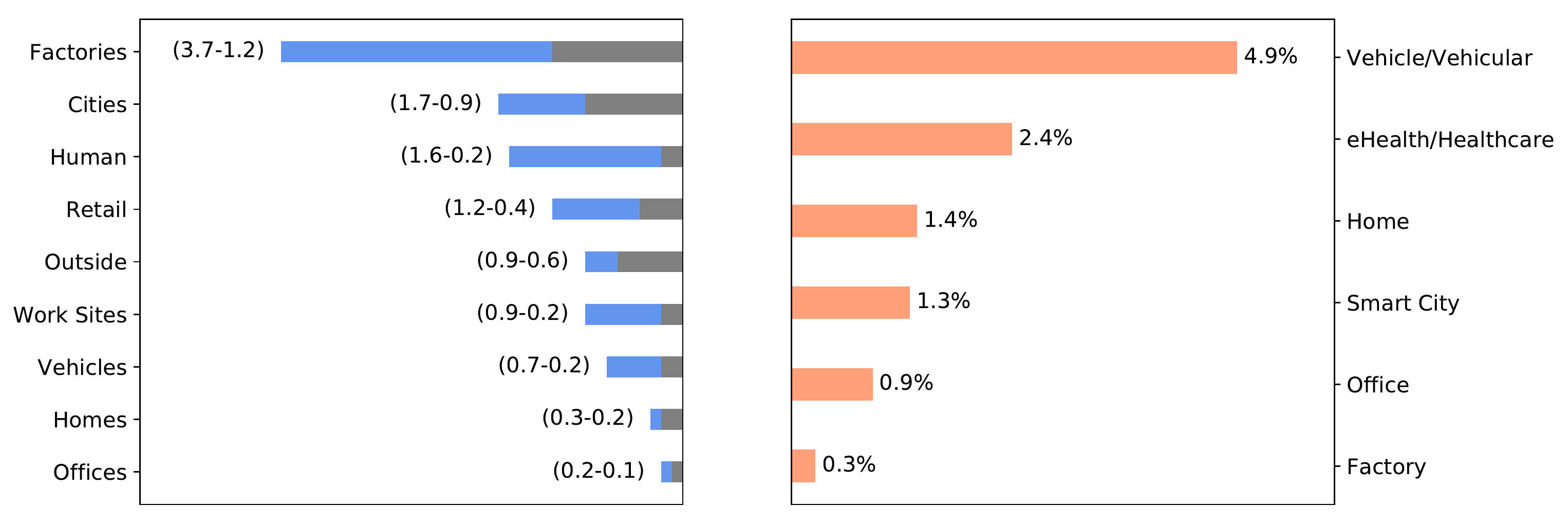}
  
  \caption{(Left) Potential economic impact (in \$ trillion) of the IoT in different sectors (high estimate--low estimate) according to \cite{McKinsey2015} vs. (Right) percentage of the selected research papers in our corpus with the listed verticals in the abstract.}
  \label{fig:mckinsey-papers}
\end{figure*}

One could argue that the manufacturing sector is just a small fraction of the potential users of wireless technologies.
However, as {shown} in \fig{fig:mckinsey-papers} (left), this industry is the sector with the highest IoT economic potential, considerably higher than that of smart-cities and transportation systems \cite{Verizon2017}.
We compare this potential impact with the percentage of papers that include any wireless-related term in their abstracts
and  one of the tokens \emph{vehicle/vehicular}, \emph{ehealth/healthcare}, \emph{home}, \emph{smart-city/cities}, \emph{office} or \emph{factory/factories} in \fig{fig:mckinsey-papers} (right). 
First, note that the articles mentioning any of these sectors amounts to $11.2\%$ of the selected papers in our corpus. 
Thus, in general, research seems to be not explicitly focused on verticals, with, perhaps, the exception of the telecommunications sector.
Second, we note that the high expected economic impact on factories is not accompanied by proportionate research attention.
Papers including \emph{factory/factories} tokens in the abstract amount to only $0.3\%$ of the total.

\subsection{Undervalued industrial interests}

We analyze now in more detail industrial concerns which are far from academic research trends.
These are mainly related to the long lifespan of industrial machines (which may exceed 20 years) as well as the typically long payback periods (around 10 years). 
This has serious implications, 
as it is necessary to fit in the current technological context machinery that appeared even before the Internet was born,
{as well as to devise a clear roadmap for the candidate technologies.} 

\subsubsection*{Retrofitting}
Legacy equipment is the main pillar of most of today's factories. 
Retrofitting, 
involves using IoT-ready connectivity solutions 
that extend the capabilities of legacy equipment.
Protocol conversion is key in this context in order to enable communication between the legacy protocols used by the equipment's components 
and modern assets that rely on Internet-based connectivity. 
However, this must-have aspect has attracted little academic attention in recent years. 
For instance, we found only 130 articles (1.4\% of the total) in which at least one of the following word stems appear: \emph{retro\texttildelow, obsole\texttildelow, longlas\texttildelow} or \emph{legac\texttildelow} 
(all of which can be associated to the concept \emph{retrofitting}).
This contrasts with the vision of the industry (\fig{fig:iiot-morgan}), 
in which the legacy-installed base is among the top 3 concerns.

\subsubsection*{Interoperability}

Machine interoperability has important open issues. 
Legacy equipment was not designed to communicate with other devices and systems. 
The proprietary nature of legacy protocols was even seen as a method for market positioning.
This makes the understanding between multi-vendor equipment complex. 
This issue remained latent for years in the isolated production lines,
but it can emerge now that the smart factory vision requires a high degree of interoperability. 
This partly explains 
why the industry considers standardization to be among the most important challenges 
to IIoT adoption (\fig{fig:iiot-morgan}).
Yet, it seems that this problem has attracted little interest among the academic community. 
For example, stems related to the concept \emph{interoperability} (such as \emph{interop\texttildelow, fragmen\texttildelow, compatib\texttildelow} and \emph{compli\texttildelow}) appear in 235 abstracts (2.6\% of the sample) and we have to bear in mind that they often appear associated with other concepts, such as \emph{packet} instead of \emph{market fragmentation}.

\subsubsection*{Maintainability}
{%
Reliability is perhaps the main preoccupation in the manufacturing industry. 
However, when industrialists use this term, they are usually expressing a 
a broader concern, 
beyond the concept specifically tied to reliable communications. 
Maintenance is a cross-dimensional discipline that 
is implicitly addressed
in most of the topics discussed previously.
For example, preventive maintenance is an active effort to improve reliability at the system level.
In turn, once a failure occurs, effective and fast repair is required. 
From this perspective, i.e. corrective maintenance, the need for standardization arises naturally. 
It also justifies the fear of fragmentation, 
raising the value of compatibility
for mitigating long-term concerns about equipment.
 
Despite its connection with all these important topics, 
the term \textit{maintenance} appears in less than 0.8\% of the articles analyzed. 
Furthermore, in many cases it is associated with the replacement of batteries which, 
as we will see next, 
are rarely found inside a factory. 
A concept commonly used in telecommunications is maintainability, which
measures the expected time to recover the operating state of a system. 
It is therefore an indicator 
that aggregates many of the concerns related to maintenance. 
Notably, this term appears only once.
}


\subsection{Unfitted research interests}

We now focus on the aspects to which academia dedicates considerable research efforts but that, according to \fig{fig:iiot-morgan} and our interviews, are considered irrelevant in the current digital transformation.
We argue that despite the potential of the IIoT in this industry, 
academia is driven by other interests.%

\subsubsection*{Energy efficiency}

Perhaps one of the most surprising mismatches identified in our interviews 
is the apparent lack of concern for low-power communication.
This is also confirmed in \fig{fig:iiot-morgan}, where power consumption does not even appear among the main challenges for IIoT adoption.
This is not the case in many other contexts. Energy efficiency is important in cellular communications, wearables and, in general, in any battery-powered device. 
Even in the IoT, applications such as agriculture monitoring and utility metering are sensitive to energy consumption.
However, mostly all factories have access to an electrical outlet everywhere
and the energy required for communications is negligible compared to the total electrical supply in a manufacturing plant.
The important aspect here is that protocols targeting low-power wireless communications intrinsically entail a trade-off between consumption and reliability, while the latter is key to this sector. 

However, energy efficiency is indeed gathering considerable research attention from academia, most probably motivated by the other many contexts
in which energy consumption is pivotal. 
Among the 9,011 articles in our dataset,
we found 1,697 abstracts (18.8\%) with one or more of the following stems:
\emph{batter\texttildelow, lowpow\texttildelow, energyharvest\texttildelow, energyeffi\texttildelow, lowener\texttildelow}, which indicates that the focus from academia on wireless energy efficiency is huge.

\subsubsection*{Wireless sensor networks heritage}

The Wireless Sensor Networks (WSN) literature is vast. %
14.2\% of the articles published in the last 3 years in the selected conferences and journals include \emph{Wireless Sensor Networks} as a keyword (\fig{fig:iiot-morgan}).
The impact of WSNs in the real world 
is out of the scope of this article.
What is clear, though, is that the research community has been greatly attracted 
to WSN application scenarios, which are distant from the industrial landscape.
Additionally, many researchers have evolved from WSN research to the IIoT. 
However, although the profile of the IIoT and WSN researchers may be similar, the two topics 
do not have so much in common. 

This progression of WSN researchers to the IIoT sphere results in proposals for the IIoT that are reasonable for WSNs but that are far from industrial requirements. %
Energy efficiency, aspect already described above, can also be considered part of the WSN legacy.
Another illustrative example is \emph{compressive sensing} (66 matches in our database): 
While exploiting spatial and temporal correlations might make sense in the typical scenarios envisioned by the WSN community (such as environmental monitoring),
it becomes inapplicable in an industrial setting.

\subsubsection*{Research trends}

The academic community is constantly seduced by new research trends.
For instance, consider the 
recently increasing attention in single-hop, long-range communications 
(110 abstracts contain terms such as \emph{lora\texttildelow, singlehop\texttildelow{} and longrange\texttildelow)}.
This focus reflects the needs of certain IoT applications (e.g. utility metering), 
which in recent years have been addressed by technologies such as SigFox and LoRa.
This situation reminds us of the great research effort carried out in multi-hop technologies  
influenced by the expectations created by WSN and encouraged by the success in sectors such as oil and gas
(we still found 239 abstracts containing \emph{multihop\texttildelow} in the 2015-2017 period).
In that case the main driver was energy optimization 
and currently it is reaching the Internet at long distances (still under strict energy constraints).
However, the focus on these and other hot topics has left little room for more suitable technologies to this industry.

\section{Convergence Paths}
\label{sec:paths}

We could imagine the heavy machinery, characteristic of the manufacturing industry, as the structure of an old building that we are compelled to preserve.
From this perspective, wireless adoption resembles a 
restoration task: reinforcing the structure (that is, machinery) and building on top of existing elements.
We next propose some paths to materialize this vision as ways for academia to move forward.

\subsection{Added value in the current machinery}

Factories are designed for easy access.
The hard-to-reach places are found in the depths of the machinery: 
sensors in spots that require high insulation, parts with complex 3D movements, rotating heads, etc. 
Machines are full of connecting tubes that wrap signal buses. 
The rigidity of these hoses, especially those that enclose fiber optic connections,
hinder and slow down the movements of internal parts. 
Wire-replacement 
can give machines more freedom for 3D movements and faster mechanics. 
Collaboration with machine vendors is essential
for understanding these mechanical limitations and offering solutions compatible with existing equipment. 
Forget about low-power, wire-replacement requires performance and reliability to achieve links comparable to the technologies being replaced.

One solution for addressing wire replacement in machinery can be mmWave. 
mmWave is the only technology at this moment able to provide fiber-optic performance and, thus, deliver the required bandwidth 
to the most demanding industrial applications.
In this setting, the distances to cover are small and even if no line-of-sight is possible, there is the possibility to take advantage of reflective surfaces, which are common in an industrial context.
We believe this is a new research area that may potentially help close the current academia-industrial gap.
It will involve the study of mmWave physical channel measurements, models of the industrial setting and evaluation of the performance at higher layers of the protocol stack, with a special emphasis on reliability  \cite{cheffena2016industrial}.

\subsection{New opportunities}

Wireless communication is particularly useful when the agents are mobile. 
In the manufacturing industry, we can find several examples of mobile targets, ranging from tools and auxiliary machinery (like forklifts) to workers and items being produced.
The heterogeneity and mobile nature of the assets involved make this topic a new opportunity for wireless. 

To make this a reality, researchers must work in collaboration with the industrial counterpart, defining realistic specifications in terms of required bandwidth, latency and reliability. 
Based on these requirements the use of licensed-band technologies will most probably be required in order to meet the specifications.
This may be achieved by forthcoming 5G ultra-reliable low-latency services \cite{Huang2018} 
and by renting resources to network operators via, for instance, slicing and/or private LTE networks.
This prospective research topic includes evaluating network performance as well as economic incentives, business models and novel service level agreements.

\subsection{Interoperability and backward/forward-compatibility}

Plug-in technologies must maintain interoperability with legacy interfaces and eventually bridge to new or evolved standards. 
The academia and standardization bodies should look at the long-term coexistence of legacy and recent technologies and define mechanisms to ensure \textit{forward-compatible} specifications, 
robust to a multi-decade evolution. {These can take advantage of the directions taken by the IETF to promote the full deployment of IPv6.
Moreover, the seamless interoperability proposed in the design of 5G (e.g extended numerology in NR) can be used as a reference model. If this model becomes fully adopted in the industry, it may assist towards long-lasting equipment as its extensible design can facilitate coexistence and the upgradability of the installed base.}

In addition, despite having seen in the last 10 years
an interesting growth 
in Software Defined Radios (SDR), we do not perceive an obvious use of the technology to achieve future-proof and universal interoperability 
(interestingly, \emph{forward-compatibility} does not even appear in the analyzed corpus).
Research lines in that direction may pave the way to quick adoption of new wireless standards as a simple firmware update. 

\subsection{Materializing security}

Academia is already focusing on security in wireless communications.
Following the analysis presented in the previous section, we found that
the stems: 
\emph{secrec\texttildelow, privac\texttildelow, eavesdrop\texttildelow, jamm\texttildelow, authe\texttildelow}
appear in 459 abstracts (5.1\% of the total).
We believe these efforts should continue and materialize into products in the years to come.
An important aspect to note is that 
highly-directional communications, like those mentioned earlier, also offer advantages in terms of security,
as these links are less prone to security attacks compared to omnidirectional transmissions.

\subsection{Demonstrating reliable wireless communications}

{Wireless research is largely devoted to improving wireless reliability. 
Indeed, in the analysis presented above, we found that stems related to reliability 
(\emph{qos, qualityof\texttildelow} and \emph{redundan\texttildelow}) appear in 696 abstracts (7.7\% of the total).
Indeed, as we have previously seen, reliability is imperative for the industry.
However, wireless technologies have yet to demonstrate the reliability and guaranteed latency required by industrial applications.}

\section{Final Remarks}
\label{sec:conclusion}

The needs of the manufacturing industry are characterized mainly 
by the fact that the equipment used is built to last for several decades.
While concerns such as reliability and security are accompanied by considerable research efforts, aspects related to long-lasting machinery
-- such as retrofitting, interoperability and maintainability --
seem to not be among the most important academic interests.
Wireless research is instead largely influenced by other design considerations 
such as energy efficiency, and it is affected by the WSN heritage.
As a consequence, the common requirements and settings of proposed wireless solutions are distant from the needs of this industry.

We propose directions for change 
that aim to contribute to an effective wireless revolution in the manufacturing industry in coming years, 
thus helping to materialize the industry 4.0 concepts.
These include continued efforts toward: 
reliability and security; 
novel research on high-frequency and directional technologies such as mmWave;
private/rented cellular 5G networks;
and virtualizable radio equipment.
All of these 
should aim to add value to existing equipment, 
devising new opportunities and even 
revisiting unsolved problems 
that may no longer qualify as hot topics.


\bibliographystyle{IEEEtran}
\bibliography{commag-paper}

\vfill
\newpage

\begin{IEEEbiographynophoto}{Borja Martinez}
received his B.Sc. in physics, the M.Sc. degree in microelectronics 
and the Ph.D. in informatics from the Universidad Aut\'onoma de Barcelona (UAB), Spain, 
where he was assistant professor from 2005 to 2015, 
combining this activity with applied research in the private sector. 
He is currently a research fellow at the IN3-UOC. 
His research interests include low-power wireless technologies, energy management policies and algorithms. 
\end{IEEEbiographynophoto}

\begin{IEEEbiographynophoto}{Cristina Cano}
holds a Ph.D. degree in information, communication, and audiovisual media technologies from Universitat Pompeu Fabra in 2011. 
She has been a research fellow at the Hamilton Institute of the National University of Ireland, Trinity College Dublin and Inria. 
She is now an associate professor at UOC. 
Her research interests include the coexistence of wireless networks, distributed resource allocation, and online optimization. 
\end{IEEEbiographynophoto}

\begin{IEEEbiographynophoto}{Xavier Vilajosana}
received his B.Sc. and M.Sc in Computer Science from Universitat Polit\`ecnica de Catalunya 
and his Ph.D. in Computer Science from the Universitat Oberta de Catalunya. 
He has been researcher in the area of industrial communication technologies for more than 15 years, 
at Orange Labs, at UC Berkeley and lately at UOC. 
His research interests include industrial wireless communication technologies and standardization.
\end{IEEEbiographynophoto}


\vfill

\end{document}